\documentclass[journal=jacsat,manuscript=article]{achemso}
\setkeys{acs}{keywords = true}
\usepackage[version=3]{mhchem} 

\pdfoutput=1
\usepackage{siunitx}
\usepackage{enumerate}
\usepackage{lineno}
\usepackage[capitalise]{cleveref}
\usepackage{braket}
\usepackage{comment}
\usepackage{booktabs}
\usepackage{array}
\usepackage{graphicx}

\newcommand\ps{P\nobreakdash{-}SNSPD}

\author{Lorenzo Stasi}
\affiliation[ID Quantique SA]
{ID Quantique SA, Rue Eugène-Marziano 25, CH-1227, Genève, Switzerland}
\alsoaffiliation[Univeristy of Geneva]
{Group of Applied Physics, Univeristy of Geneva, Rue de l'Ecole-de-Médecine 20, CH-1211, Genève, Switzerland}
\altaffiliation{These authors contributed equally}

\author{Towsif Taher}
\email{towsif.taher@unige.ch}
\altaffiliation{These authors contributed equally}
\affiliation[Univeristy of Geneva]
{Group of Applied Physics, Univeristy of Geneva, Rue de l'Ecole-de-Médecine 20, CH-1211, Genève, Switzerland}

\author{Giovanni V. Resta}
\email{giovanni.resta@idquantique.com}
\affiliation[ID Quantique SA]
{ID Quantique SA, Rue Eugène-Marziano 25, CH-1227, Genève, Switzerland}

\author{Hugo Zbinden}
\affiliation[Univeristy of Geneva]
{Group of Applied Physics, Univeristy of Geneva, Rue de l'Ecole-de-Médecine 20, CH-1211, Genève, Switzerland}

\author{Rob Thew}
\affiliation[Univeristy of Geneva]
{Group of Applied Physics, Univeristy of Geneva, Rue de l'Ecole-de-Médecine 20, CH-1211, Genève, Switzerland}

\author{Félix Bussières}
\affiliation[ID Quantique SA]
{ID Quantique SA, Rue Eugène-Marziano 25, CH-1227, Genève, Switzerland}

\title{Enhanced detection rate and high photon\nobreakdash{-}number efficiencies with a scalable parallel SNSPD}

\keywords{SNSPDs, single-photon detectors, parallel architecture, ultra-fast detection, photon-number resolution, quantum communication, quantum computing}

\begin{document}

\begin{abstract} 
Since their inception, superconducting nanowire single\nobreakdash{-}photon detectors have been enabling quantum optical applications and the rise of the photonic quantum industry. The evolution in the detector design and read\nobreakdash{-}out strategies has led to the introduction of devices with a plurality of independent pixels, which have been able to operate with high system detection efficiency at high speed while also supporting photon number resolution capabilities. However, this comes at the cost of a complex readout that requires one coaxial cable for each pixel of the array. Here, we report a 28\nobreakdash{-}pixel SNSPD with a dedicated parallel architecture that, while maintaining a simple readout with a single coaxial line, enables the detector to operate at high speed with low-performance degradation. The device shows a maximum single-photon efficiency of 88\% and is able to maintain its efficiency above 50\%, coupled with a timing jitter lower than 80~ps, up to a detection rate of 200 million counts per second. The detector also provides state-of-the-art photon-number-resolving performances with a 2-photon efficiency of 75\% and a 3-photon efficiency of 62\%.
\end{abstract}

\section{Introduction}

Single photons can be employed in applications ranging from quantum communication to photonic quantum computing and quantum-enhanced imaging and sensing \cite{natarajan2012superconducting}. The foreseen development of optical quantum computers and quantum networks will require generation, manipulation, and detection of quantum light, and in this context, distribution of entanglement using single-photons as flying qubits is paramount \cite{slussarenko2019photonic, azuma2023quantum}. The unprecedented capabilities that would result from the deployment of these technologies have driven the research community to push the limits of the performances for the fundamental components of such systems, namely single-photon sources~\cite{somaschi2016near, meyer2020single, tomm2021bright}, photonic integrated circuits~\cite{wang2020integrated, wang2024lithium}, quantum memories~\cite{simon2010quantum} and single photon detectors~\cite{hadfield2009single, lita2022development}. In the case of single-photon detectors, near-unitary efficiency, few dark counts, low jitter, high detection rates, and photon\nobreakdash{-}number resolution (PNR) will be required. In particular, the ability to resolve photon\nobreakdash{-}number states is an essential functionality for different applications such 
as quantum repeater protocols~\cite{sangouard2011quantum}, generation and distribution of heralded multi\nobreakdash{-}photon entanglement~\cite{Cao2024GHZ, marcellino2024toward, gerrits2010generation},
improved heralded single-photon sources~\cite{stasi2023high, davis2022improved, sempere2022reducing}, gaussian boson sampling~\cite{arrazola2021quantum, madsen2022quantum}, linear optical quantum computing~\cite{knill2001scheme, bartolucci2023fusion, maring2024versatile} and enhanced quantum metrology~\cite{von2019quantum, schmidt2018photon}. Without a scalable solution for a device that can perform single-photon detection and PNR discrimination with high $n$\nobreakdash{-}photon efficiencies \cite{bienfang2023single} at high speed, QKD protocols, quantum networks, and optical quantum computers will suffer from severe speed limitations, hindering these technologies from achieving their full potential.

Amongst the wide variety of single-photon detectors, single-pixel superconducting nanowire single\nobreakdash{-}photon detectors (SNSPDs) \cite{gol2001picosecond} have demonstrated remarkable results in terms of system detection efficiency (SDE)~\cite{reddy2020superconducting, chang2021detecting}, dark counts~\cite{shibata2015ultimate}, timing jitter\cite{korzh2020demonstration, esmaeil2020efficient} and capability to work at a wide range of wavelengths \cite{wollman2017uv, colangelo2022large}. Crucially, in their single-pixel implementation, these nanowires are limited by their recovery time of a few tens of nanoseconds \cite{esmaeil2017single, autebert2020direct} and cannot detect single-photon events faster than a few million counts per second (Mcps), without a sharp degradation of their efficiency and timing jitter \cite{mueller2023time}. The degradation of SDE and jitter drastically impacts the intrinsic PNR capability of such single-pixel SNSPDs, as low jitter and high SDE are required in order to have high $n$\nobreakdash{-}photon efficiencies \cite{cahall2017multi, zhu2020resolving, sauer2023resolving, schapeler2023well}. Another limitation to the use of single-pixel SNSPDs is that the PNR functionality can only be used with light pulses of a few tens of picoseconds at most \cite{zhu2020resolving, sauer2023resolving, schapeler2023well}, thus preventing their use in applications with light sources that produce photons with longer pulse durations~\cite{brydges2023integrated}. 

To overcome the shortcomings of the single\nobreakdash{-}pixel design, approaches based on time\nobreakdash{-}multiplexing \cite{achilles2004photon, deng2023gaussian} or spatial\nobreakdash{-}multiplexing have been adopted~\cite{dauler2007multi, lusardi2017photon, wang2019boson, bayerbach2023bell, Cao2024GHZ}. In the time\nobreakdash{-}multiplexing approach, fiber loops and optical beam splitters are used to distribute the photons in different temporal bins, thus allowing discrimination of different photon-number states by a single-pixel SNSPD~\cite{deng2023gaussian}. This scheme suffers from the additional losses of the optical beam splitters and fiber loops and is inherently slower than other approaches. Spatial multiplexing can be exploited either by using multiple individual detectors and splitting the light with optical beam-splitters \cite{lusardi2017photon, wang2019boson, bayerbach2023bell, Cao2024GHZ} or by developing dedicated SNSPD designs, where the detection area is divided into several smaller SNSPDs (called pixels) that are biased and read-out individually \cite{dauler2007multi}. The former approach does not address the speed limitation of single-pixel devices and introduces additional losses due to the use of optical beam splitters. With the latter approach, the losses of the beam-splitters are removed, and it is possible to maintain high SDE at higher count rates and even reach maximum detection rates in the order of Giga-cps (Gcps) \cite{zhang201916, craiciu2023high, resta2023gigahertz}. With these arrays, the photon-number information is encoded in the number of pixels that have clicked and a high number of pixels is necessary to minimize the probability that multiple photons are absorbed by the same pixel. However, this comes at the cost of a readout whose complexity scales linearly with the number of pixels. In fact, one coaxial cable per pixel is needed, and coincidence analysis has to be performed to extract information on the detected photon\nobreakdash{-}number state, possibly limiting real\nobreakdash{-}time operations. Crucially, the limited number of connections that can be made to the coldest stage of a cryo-cooler drastically limits the scaling of this architecture. 

A solution to these limitations has been the development of a dedicated parallel architecture that allows one to combine the signal of the pixels of the SNSPD in parallel directly on-chip, thus allowing the use of a single cryogenic cable to read out a parallel-SNSPD ({\ps})~\cite{perrenoud2021operation}. This architecture also demonstrated PNR capabilities~\cite{stasi2023fast, stasi2023high}, where the information on the detected photon\nobreakdash{-}number state is encoded in the amplitude of the signal pulse. However, the small number of active pixels was limiting the maximum achievable detection rate (6 pixels, 100~Mcps detection rate with $\sim35\%$ absolute efficiency)\cite{perrenoud2021operation} and the photon\nobreakdash{-}number efficiencies (4 pixels, 48\% for 2-photon efficiency) \cite{stasi2023fast}. 

In this work, we report the development of an optimized {\ps} architecture that allows for an increased number of pixels, thus enabling a unique combination of state-of-the-art performances and scalability. We demonstrate a 28\nobreakdash{-}pixel {\ps}, with 88\% SDE at the single photon level, that is able to reach 200~Mcps at 50\% absolute SDE. The parallel architecture is able to maintain a timing jitter below 80~ps full width at half maximum (FWHM) at a 200~Mcps detection rate while using a single coaxial line readout. We perform an in-depth analysis of the PNR performances of the device and demonstrate 2\nobreakdash{-}photon efficiency of 75\% and 3\nobreakdash{-}photon efficiency of 62\%. Finally, we compare our technology to other PNR methodologies, highlighting the strengths and shortcomings of each method in an effort to bring better clarity and understanding to a field that has seen rapid progress in recent years. 

\section{Device fabrication and characterization}

\begin{figure}
    \centering
    \includegraphics[width=\textwidth]{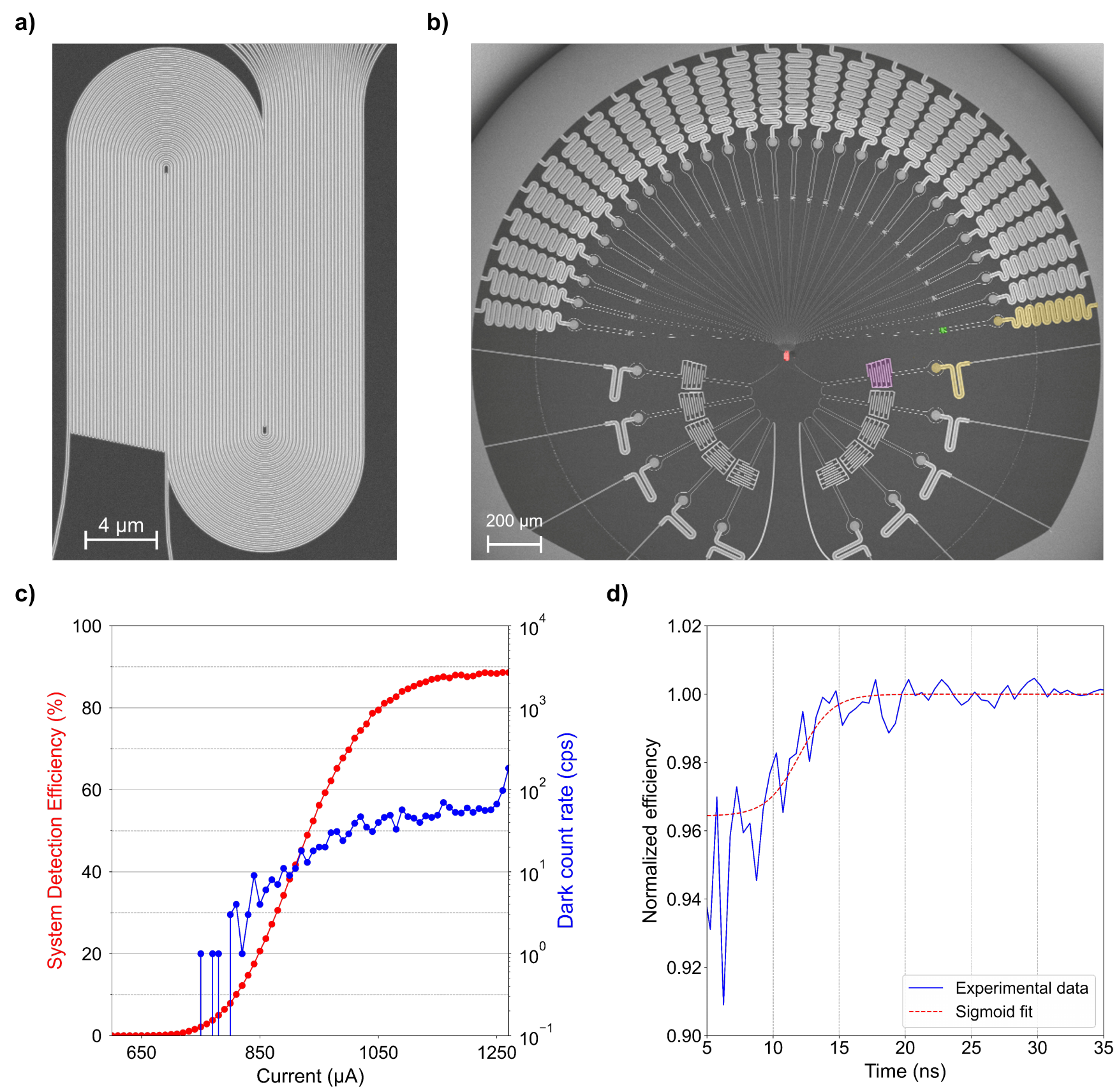}
    \caption{\textbf{a)} SEM image of the active 28\nobreakdash{-}pixels arranged in an interleaved geometry. The active area of the detectors is $\sim$\SI{225}{\um^2}. \textbf{b)} Recolored SEM image of the entire parallel architecture. In the center, the 28 active nanowires are recolored in red. In the top half, there are meandered series resistors (yellow), ensuring a uniform bias of the pixels, and small series inductors (green), which tune the kinetic inductance of the pixels. In the bottom half, there are the additional unexposed micro-wires (purple) and resistors (yellow) needed to avoid electrical cross-talk due to the current redistribution effect. \textbf{c)} System detection efficiency (red circles) and dark-count rate (blue circles) of the 28\nobreakdash{-}pixels {\ps} measured with a 1550~nm CW laser at a photon flux of {$\sim$$10^5$} photon/s. \textbf{d)} Recovery of the detector efficiency as a function of time. The efficiency reaches 96\% of its maximum value within 5~ns of a previous detection and fully recovers in less than 15~ns.}
    \label{fig:SEM_SDE_RT}
\end{figure}

The {\ps} consists of 28 nanowires fabricated on ${\sim}9$~nm thick NbTiN layer (see supporting information for a detailed description of the fabrication process). The nanowires are $\sim$\SI{100}{\nm} wide with $\sim$\SI{100}{\nm} spacing, arranged in an interleaved geometry to ensure uniform illumination of all pixels (\cref{fig:SEM_SDE_RT}a). 
The active pixels are connected in parallel on\nobreakdash{-}chip and read out with a single coaxial cable, as shown in \cref{fig:SEM_SDE_RT}b. In the top half of the image, there are 28 meandered series resistors (yellow), that ensure a uniform bias of each active pixels. The optimal value of these resistors depends on the kinetic inductance of the pixels and needs to be chosen in order to avoid any possible latching at high count-rates. The value in this demonstration is 20~Ohm. Small series inductors (green) are also present, which tune the kinetic inductance of the pixels. To avoid electrical cross-talk between the pixels due to the current redistribution effect after a photon\nobreakdash{-}detection event, additional wider micro-wires (purple) and resistors (yellow) are connected in parallel to the active 28\nobreakdash{-}pixels outside the illumination area of the fiber~\cite{perrenoud2021operation}. The values of the inductances of the unexposed micro-wires (${\sim}30$~nH) and of the resistors (${\sim}5$~Ohm) are chosen to enable the simultaneous detection of up to 8 photons without any electrical crosstalk between the nanowires. With respect to previous demonstrations of {\ps} \cite{perrenoud2021operation,stasi2023fast}, the number of active pixels has been drastically increased (from 6 and 4 to 28). Moreover, we transitioned from adjacent pixels, separated by a gap of 800~nm, that was needed to avoid thermal crosstalk between the pixels \cite{perrenoud2021operation}, to fully interleaved pixels with only 100~nm separation. These changes were made possible by the use of cristalline NbTiN superconducting material, that, thanks to its higher critical temperature and lower kinetic inductance, is less prone to thermal crosstlak with respect to amorphous MoSi used in the previous demonstrations (see supporting information for details on thermal crosstalk). With this key improvements we are able to maximize the efficiency, photon-number resolution capability and counting rate of the device. 
The detector is integrated into an optical cavity designed for maximizing photon absorption at 1550~nm (see supporting information for a detailed description of the fabrication process), coupled to an SMF-28 fiber with a self-aligning scheme \cite{miller2011compact}, and placed in a closed cycle 3\nobreakdash{-}stage cryostat (ID281, ID Quantique) at 0.8~K. We use cryogenic AC-coupled amplifiers placed at the 40~K stage and amplify the signal again at room temperature with low-noise amplifiers. We characterize the performance of the detector in terms of system detection efficiency (SDE), recovery time, count rate, and jitter. 

The SDE as a function of the bias current is measured at 1550~nm by illuminating the {\ps} with a CW laser (Thorlabs MCLS1) at a photon flux of $10^{5}$ photon/s (see \cref{fig:SEM_SDE_RT}c). The {\ps} shows a long plateau region of constant SDE, reaching a maximum SDE of 88\% with only $\sim$60 dark counts/s. \cref{fig:SEM_SDE_RT}d shows the normalized SDE as a function of time. The high number of active pixels (28) of our designs allows the efficiency to only drop by a value of $\mathrm{1/28^{th}}$ after a single-photon detection, and their lower kinetic inductance allows for a fast recovery of the SDE. The SDE remains 96\% of its maximum value after the detection of a single\nobreakdash{-}photon, and after 15~ns it fully recovers. The shorter length of the individual nanowires, coupled with the presence of the plateau, helps us to select an operation current for the device that exhibits negligible thermal crosstalk between the pixels (see supporting information for details on thermal crosstalk analysis). The results of our crosstalk analysis on the {\ps} are consistent with previous reports in the literature on independent multi-pixel SNSPD \cite{resta2023gigahertz, craiciu2023high}, highlighting the importance of short pixels with a reduced kinetic inductance in order to minimize thermal crosstalk.

\begin{figure}
    \centering
    \includegraphics[width=\textwidth]{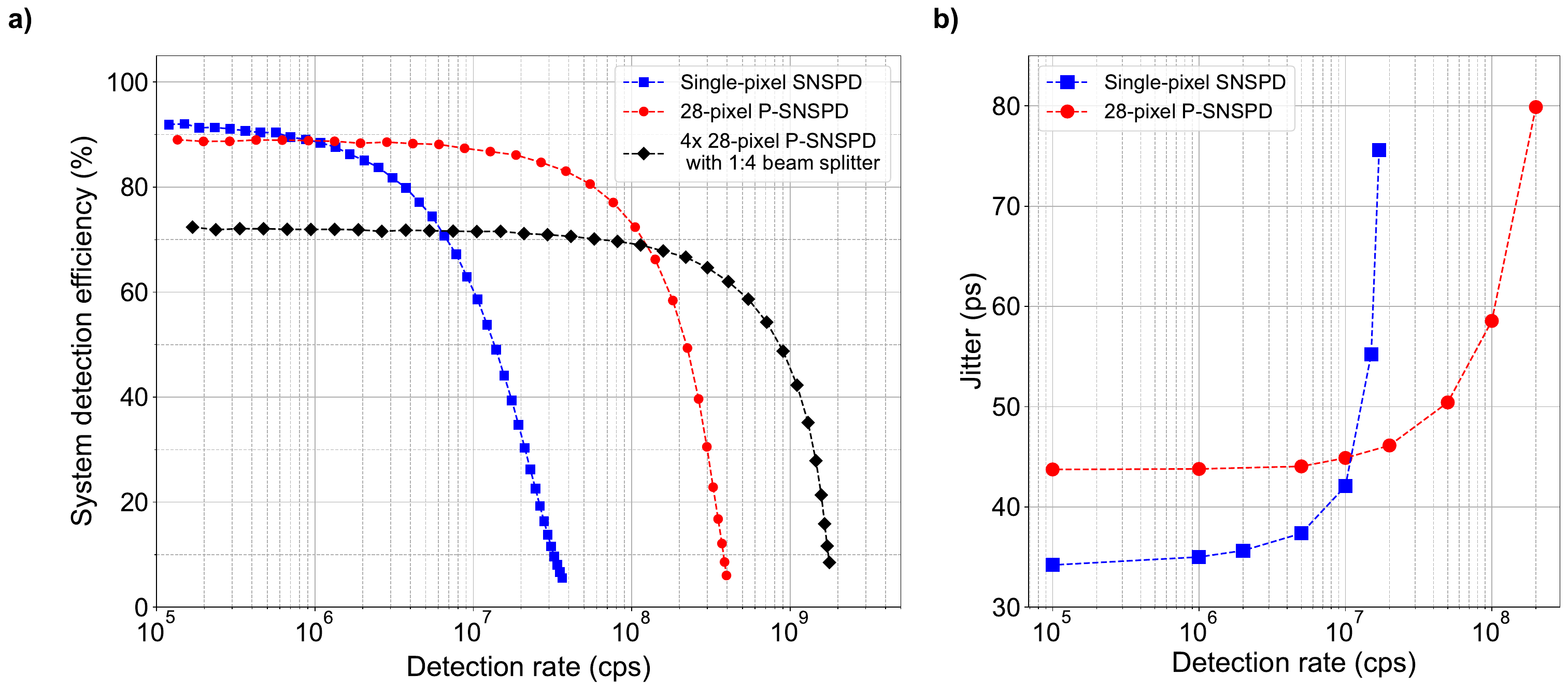}
    \caption{\textbf{a)} SDE as a function of the detection rate measured with 1550~nm CW laser for a 28-pixel {\ps} (red circles), a standard single-pixel SNSPD (blue squares) and 4x 28\nobreakdash{-}pixel {\ps} with optical beam-splitter (black diamonds). Note that for the 4x 28\nobreakdash{-}pixel {\ps}, we used off-the-shelf components and did not factor out any losses. \textbf{b)} Timing jitter as a function of the detection rate for a 28\nobreakdash{-}pixel {\ps} (red circles) and for a standard single-pixel SNSPD (blue squares).}
    \label{fig:CR_Jitt}
\end{figure}

In \cref{fig:CR_Jitt}a, we highlight a key advantage demonstrated by this architecture where a high detection rate is achieved by the {\ps} using only a single coaxial line read-out. We compare the evolution of the SDE of the {\ps} as a function of detection rate with a conventional single-pixel SNSPD fabricated on the same wafer. For the {\ps} detector, SDE remains above 80\% up to 50~Mcps and above 50\% up to 200~Mcps. We define the maximum count rate (MCR) as the detection rate at which the normalized efficiency decreases by 3~dB\cite{craiciu2023high}. The detector demonstrates an MCR of ${\sim}$250~Mcps, showing more than a 10-fold improvement with respect to the conventional single-pixel SNSPD that has an MCR of only ${\sim}$20~Mcps.

We further study the suitability of our detector for applications in high-rate quantum key distribution (QKD), where detectors with Gcps detection rates are required to achieve high ($\sim$100~MHz) secret key rates. Such high detection rates have previously been realized using spatially multiplexed SNSPDs with independent biasing and read-out of each pixel as demonstrated in Ref. \cite{li2023high, grunenfelder2023fast}, where detectors requiring 16 and 14 cryogenic read-out channels were used respectively. In \cref{fig:CR_Jitt}a, we demonstrate Gcps detection rates using four {\ps} detectors requiring only four coaxial lines. To achieve this, we split the optical input using a commercially available 1:4 optical beam splitter and send the outputs of the splitter to four detectors with 88\%, 86\%, 85\%, and 82\% SDE, respectively. Each {\ps} is read out using one coaxial line (see supporting information for details of the measurement setup). We achieve 1~Gcps detection rate at $\sim$45\% absolute SDE and a MCR of $\sim$1.3~Gcps, with a much simpler read\nobreakdash{-}out scheme compared to Ref.~\cite{zhang201916, craiciu2023high, resta2023gigahertz,li2023high}. With this method, achieving such extreme detection rates comes at the cost of losses that are introduced in the optical set-up by the 1:4 splitter (0.5~dB loss from manufacturer), connectors (0.1~dB), and mating sleeves (0.1~dB) limiting the maximum achievable SDE to $\sim$72\%. We expect that by carefully selecting an ultra-low loss splitter and splicing the fibers, it would be possible to achieve $>80\%$ maximum SDE and $>50\%$ absolute SDE at 1~Gcps detection rate. 

Another key metric for SNSPD performance is the timing jitter, which characterizes the uncertainty in the measured arrival time of a photon. A unique feature of this architecture is the capability to maintain low jitter with increasing count rate without employing any jitter correction methods such as differential readout ~\cite{Colangelo2023differential} or time\nobreakdash{-}walk correction~\cite{mueller2023time}. To demonstrate this feature, we characterized the evolution of the detector jitter with respect to the detection rate by superimposing photons from a pulsed laser (NuPhoton Technologies, 27~MHz repetition rate, 6~ps pulse width) and a CW laser (Thorlabs, MCLS1) via a 50:50 beam splitter. The photons from the CW laser were used to achieve high detection rates, and the pulsed laser provided photons to characterize the jitter (see supporting information for details of the measurement setup). \cref{fig:CR_Jitt}b shows results from this characterization for a {\ps} detector and the same conventional single-pixel detector as before. Both detectors use the same read-out scheme with cryogenic amplification at 40~K and further amplification at room-temperature. As can be seen, the {\ps} presents a slightly elevated jitter at low count-rates compared to the single-pixel SNSPD ($\sim$43~ps FWHM and $\sim$34~ps FWHM respectively). This is caused by the lower signal amplitude of the 1-click level with respect to the one of a single-pixel SNSPD, due to the current redistribution effect. In our design we optimized the values of the electrical components of the P-SNSPD architecture (series inductances and resistances of active pixels and inductance and resistance of unexposed nanowires) in order to find a balance between signal amplitude and mitigation of current redistribution. Further improvements of the {\ps} architecture or of the amplification chain could mitigate this effect. However, when looking at the jitter at high count rates, the advantage of having several pixels becomes evident. For our {\ps}, the jitter remains below 60~ps FWHM up to 100~Mcps and below 80~ps FWHM until 200~Mcps, whereas, for the single pixel SNSPD, the jitter goes above 70~ps FWHM after 15 Mcps count rate.

\section{Photon-number-resolution}

In this section, we perform an in-depth study of the PNR performances of the detector. We further discuss the strengths and weaknesses of different photon number resolution methods, highlighting some key metrics that are of central importance for their operation.

\subsection{Characterization of photon-number-resolution performance}

\begin{figure}[h]
        \includegraphics[width=\textwidth]{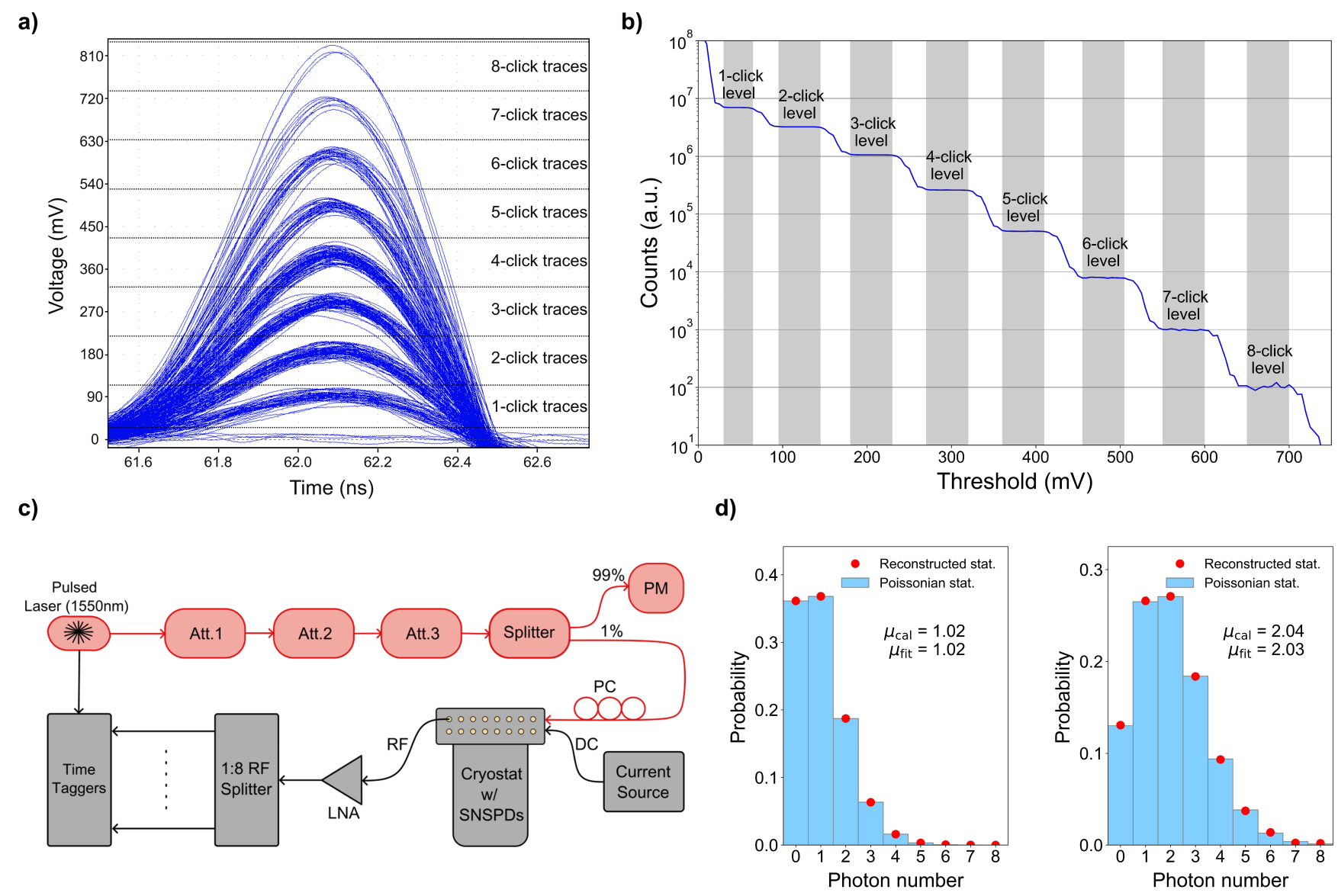}
    \caption{\textbf{a)} Electrical voltage pulse generated by the {\ps}. \textbf{b)} Plot of counts as a function of threshold, where 8 well-separated regions can be identified. \textbf{c)} Schematic of the experimental setup to acquire the photon counting statistics (PM: Power meter; PC: Polarization controller; Att.: Optical attenuator). \textbf{d)} Coherent state reconstruction for $\mu$ equal to 1 and 2, performed with a 1550~nm pulsed laser at a rate of 40~MHz.}
    \label{fig:pnr}
\end{figure}

In the {\ps} architecture, the individual pixels are electrically connected in parallel on-chip, and only a single coaxial line is needed to read out all pixels. This unique feature allows the information on the number of pixels that clicked to be encoded in the voltage amplitude of the output signal, as shown in \cref{fig:pnr}a. Since the output signal is, to a first approximation, the sum of the contribution of each pixel, the amplitude of the \textit{n-}click trace is roughly n times the 1-photon amplitude. However, this mechanism only works if the photons are arriving on the detector plane within the rise time of the electrical signal ($\sim$300~ps). For longer light pulses, the amplitudes might sum up in a non-trivial way, and no simple discrimination based on the voltage amplitude of the output can be achieved. For pulse duration below 300~ps, the clear separation between the $n$\nobreakdash{-}click amplitudes leads to a unitary probability of assigning them to the corresponding $n$\nobreakdash{-}click event (i.e.~100\% assignment probability), as exemplified in \cref{fig:pnr}b. By assigning the threshold levels for each $n$\nobreakdash{-}click event within the shaded grey region in \cref{fig:pnr}b, the corresponding $n$\nobreakdash{-}photon counts can be extracted without any cross-talk between them, and without the use of time tagging devices with ps-level resolution. This is a great advantage compared to PNR schemes based on analysis of the rising/falling edge of a standard single\nobreakdash{-}pixel SNSPD signal, where the probability of correctly assigning $n$\nobreakdash{-}photon events rapidly decreases after 3\nobreakdash{-}photons~\cite{cahall2017multi, sauer2023resolving, schapeler2023well} and that can only be used with light pulses of a few tens of ps~\cite{zhu2020resolving, sauer2023resolving, schapeler2023well}.

To describe the multi\nobreakdash{-}photon response of the 28\nobreakdash{-}pixel {\ps}, we use the model developed in Ref.~\cite{fitch2003photon} to construct the ${P_{nm}}$ elements of the detector \textbf{P} matrix, where ${P_{nm}}$ is the probability of registering an $n$\nobreakdash{-}click event when $m$ photons are incident. Due to the low probability of dark counts, all the ${P_{nm}}$ elements with $n>m$ are set to 0; thus, $\mathbf{P}$ is an upper triangular matrix. These probabilities are necessary to link the photon\nobreakdash{-}counting statistics, ${Q_n}$, namely the probability of measuring an $n$\nobreakdash{-}click event by the PNR detector, to the incoming photon statistics, $S_m$, via ${Q_n} = \sum_{m=0}^{\infty} P_{nm} \cdot S_m$. In this formalism, the diagonal elements of the \textbf{P} matrix (the ${P_{nn}}$ elements) are known as $n$\nobreakdash{-}photon efficiencies and are calculated as $\frac{N!}{(N-n)!}\left(\frac{\eta}{N}\right)^n$, where $N$ is the number of pixels, $n$ is the number of photons and $\eta$ is the single-photon efficiency of the detector. This formula shows that when using spatial multiplexing, as it is the case for {\ps}s, the $n$\nobreakdash{-}photon efficiencies are maximized by increasing $N$ and $\eta$. With an increased number of pixels ($N \gg n$), the {\ps} approaches an intrinsic PNR behavior, where the ${P_{nn}}$ are simply given by $\eta^n$. This point will be discussed further in the following sections. 

To validate the model and to demonstrate the possibility of performing photon number resolution at a high count rate, we perform coherent state reconstruction using Poissonian light at a 40~MHz repetition rate (33~ps pulse width). A calibrated mean\nobreakdash{-}photon\nobreakdash{-}number per pulse ($\mu_{\rm{cal}}$) is sent to the detector and the ${Q_n}$ is acquired by splitting the {\ps} electrical signal into 8 copies. Each one of them is connected to a time tagging device (ID Quantique, ID1000), and the proper voltage threshold is applied to get the different 8\nobreakdash{-}click events (see \cref{fig:pnr}c for a schematic representation of the experimental set-up). From ${Q_n}$, we reconstruct ${S_m}$ and fit it to a Poissonian distribution to retrieve its mean\nobreakdash{-}photon\nobreakdash{-}number per pulse ($\mu_{\rm{fit}}$). As can be seen in \cref{fig:pnr}d, there is an excellent agreement between $\mu_{\rm{cal}}$ and $\mu_{\rm{fit}}$, thus supporting the correctness of the model and a 2\nobreakdash{-}photon efficiency (${P_{22}}$) of 75\% and a 3\nobreakdash{-}photon efficiency (${P_{33}}$) of 62\% (see supporting information for the \textbf{P} matrix). The coherent state reconstruction with 40~MHz repetition rate pulses demonstrates the capability of the {\ps} to operate as a PNR detector at high count rates, thanks to the high number of pixels and their fast recovery time. In contrast, other methods of photon number resolution such as TES (limited to kHz rates)~\cite{morais2024precisely} or PNR schemes based on single\nobreakdash{-}pixel SNSPD (limited to a few MHz rates due to rapid decline of the detector efficiency and jitter at high rates, as shown in \cref{fig:CR_Jitt}), would not have been able to operate at such high-speed, failing to properly reconstruct the input light.  

\subsection{Confidence in photon number discrimination}

\begin{figure*}
        \centering
        \includegraphics[width=\textwidth]{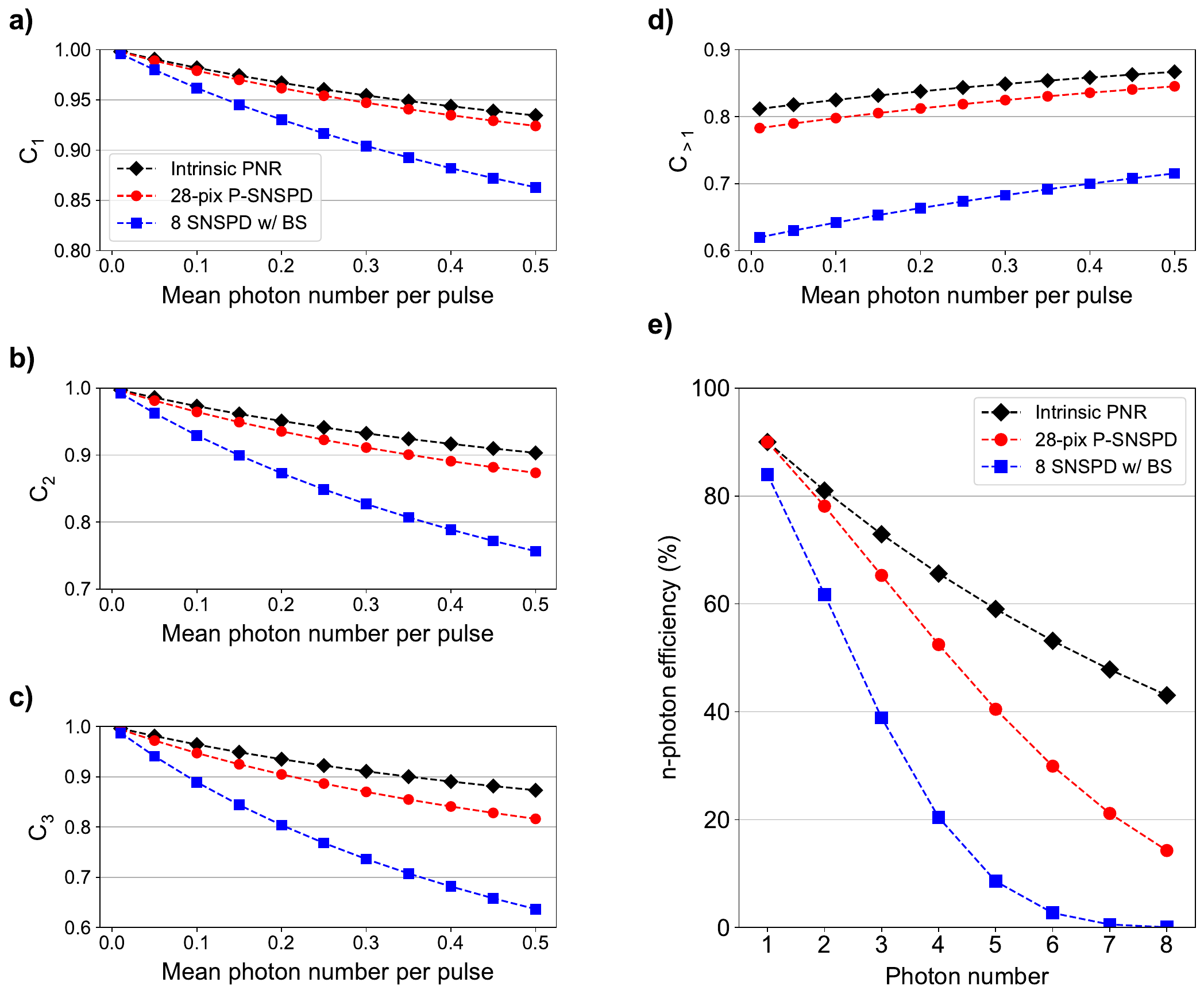}
    \caption{\textbf{a) - d)} Plot of confidence at 1 photon ($C_1$) \textbf{a}, at 2 photons ($C_2$) \textbf{b}, at 3 photons ($C_3$) \textbf{c} and at more\nobreakdash{-}than\nobreakdash{-}one\nobreakdash{-}photon ($C_{>1}$) \textbf{d} for an intrinsic PNR detector (black diamonds), a 28\nobreakdash{-}pixel {\ps} (red circles) and a hypothetical configuration involving 8 standard SNSPD with an optical beam splitter (with 0.3~dB insertion loss) (blue squares). \textbf{e)} Scaling of the \textit{n}\nobreakdash{-}photon efficiencies for the aforementioned architectures.}
    \label{fig:pnr_confidence}
\end{figure*}

In the previous section, we validated the \textbf{P} matrix model of the detector, showed high $n$\nobreakdash{-}photon efficiencies, and performed a statistical state-reconstruction of Poissonian light. Here, we investigate in more detail the response of the 28-pixel {\ps} to single-shot measurements, where at each trial, we are interested in evaluating if the detector has correctly discriminated the input photon-number state. This assessment cannot be done without placing some assumptions on the statistics of the input light, and in the following discussion, we limit our analysis to one mode of a two-mode squeezed vacuum state, which exhibits thermal statistics in the low\nobreakdash{-}$\mu$ regime. This light is generally produced by probabilistic photon\nobreakdash{-}pair sources based on spontaneous parametric down-conversion (SPDC) or spontaneous four\nobreakdash{-}wave mixing (SFWM) processes \cite{meyer2020single}. Since the \textbf{P} matrix provides information on all multi-photon detection events, it is used to answer the following question: ``What is the probability that given an $n$\nobreakdash{-}click event from the {\ps}, there were exactly $n$ photons in the input state?''. We refer to these quantities as the \textit{confidence} at $n$\nobreakdash{-}photons ($C_n$), and they can easily be calculated using Bayes' theorem:
\begin{equation}
    C_{n} = \frac{P_{nn} \cdot S_n}{Q_n} =\frac{P_{nn} \cdot S_n}{\sum_{m=n}^{\infty}P_{nm}\cdot S_m},
\end{equation}
where the denominator is the statistics of the specific $n$\nobreakdash{-}click event, and the numerator represents the subset of correct click-events for such $Q_n$. In \cref{fig:pnr_confidence}a-c, we report the confidence values at 1, 2, and 3 photons for three different PNR detection schemes: an intrinsic PNR detector (\textit{i.e.} the $n$\nobreakdash{-}photon efficiencies scale as $\eta^n$), an array composed of 8 individual single-pixel SNSPDs with splitting of optical inputs, and a 28\nobreakdash{-pixel} {\ps}. The SNSPD array with 8 individual detectors and split optical input represents a setup similar to the experiment described in Ref\cite{bayerbach2023bell} where 48 independent single-pixel SNSPDs were used to realize 6 PNR detectors (8 single-pixel SNSPD for each PNR). We perform this analysis assuming a single photon SDE of 90\% for all detectors, and for the latter configuration, we add \SI{0.3}{\dB} losses to emulate the optical beam splitter. As it can be seen from \cref{fig:pnr_confidence}a-c, for our {\ps}, the confidence values at 1, 2, and 3 photons remain respectively within 1\%, 3\%, and 7\% of the intrinsic PNR values and far outperform the 8-detector configuration.

Depending on the specific application that requires discrimination of the photon-number state, a different metric might become relevant, such as: ``What is the probability that given an input state with $>1$ photon, there will be a $>1$-click event registered?''. This probability is of great importance for any single\nobreakdash{-}photon sources that suffer from multi\nobreakdash{-}photon emissions ~\cite{stasi2023fast,davis2022improved,sempere2022reducing}, quantum communication and linear optic quantum computing protocols that rely on the capability to distinguish only between 1 and $>1$ photons~\cite{sangouard2011quantum, o2007optical, ralph2010optical}. We refer to this quantity as the \textit{confidence at more than 1 photon} $C_{>1}$. In order to answer this question, we need to re-normalize the input light statistics, considering only the subspace of the distribution where there is more than 1 photon. Given the original distribution $\mathbf{S}=[S_0, S_1, S_2, ..., S_m]$, the re-normalized statistics are:
\begin{equation}
    \mathbf{S'}=\left[0, 0, \frac{S_2}{1-S_1-S_0}, \frac{S_3}{1-S_1-S_0}, ..., \frac{S_m}{1-S_1-S_0}\right].
\end{equation}
Then, $C_{>1}$ can be expressed as:
\begin{equation}
    C_{>1} = 1- \sum_{m=2}^{\infty}P_{1m}\cdot S'_m - \sum_{m=2}^{\infty}P_{0m}\cdot S'_m ,
\end{equation}
where the second term represents the probability of registering a 1\nobreakdash{-}click when more\nobreakdash{-}than\nobreakdash{-}1\nobreakdash{-}photon is arriving on the detector, and the third term represents the probability of registering a 0\nobreakdash{-}click event when more\nobreakdash{-}than\nobreakdash{-}1\nobreakdash{-}photon is arriving on the detector. In \cref{fig:pnr_confidence}d, we show the results for the same three PNR approaches. In this case, the {\ps} remains within 3\% of the confidence values of an intrinsic PNR detector, while the 8 SNSPD configuration is considerably lower.
As hinted above, the reason behind the drastic increase in performances enabled by the 28-pixel {\ps} lies in the improved $n$\nobreakdash{-}photon efficiencies of the detector. In \cref{fig:pnr_confidence}e, we present the $n$\nobreakdash{-}photon efficiencies up to $n=8$, for the intrinsic PNR detector, the 28-pixel {\ps} and the array composed of 8 individual single-pixel SNSPDs with splitting of optical input. There is less than 3\% difference in the 2\nobreakdash{-}photon efficiency and less than 8\% difference for the 3\nobreakdash{-}photon efficiency compared to an intrinsic PNR detector, while the performances of the 8 individual single-pixel SNSPDs with splitting of optical input are far from the intrinsic level. This analysis clearly shows the advantage that stems from the use of detectors with a high number of pixels.  

\subsection{Comparison of PNR approaches}

With the following discussion, we aim to highlight some key metrics that can help to reliably evaluate and compare different schemes for photon-number resolution. Here, we define an ideal PNR detector as a device that should at least respect the following 2 properties:  
\begin{enumerate}[(i)]
    \item the $n$\nobreakdash{-}photon efficiencies should only depend on the single-photon efficiency of the detector, i.e. the diagonal elements of the \textbf{P} matrix equal to $\eta^n$ (where $\eta$ is the device efficiency and $n$ the photon number);
    \item the probability to assign each different output signal to the corresponding $n$\nobreakdash{-}click event should be unitary in the range of interest (\textit{i.e.} assignment probability of 1);
\end{enumerate}

Together with these properties, we also highlight some key features that would be required by most practical applications: 
\begin{enumerate}[(a)]
    \item absence of a strict limitation on the duration of the input light pulses;
    \item ability to work at high count rates;
    \item scalability of the read-out architecture and operational simplicity.
\end{enumerate}

We focus our discussion on the five most commonly deployed methods for PNR at telecommunication wavelengths in literature: transition\nobreakdash{-}edge sensors (TESs)~\cite{irwin1995application, lita2022development, morais2024precisely}, single-pixel SNSPDs with optical beam-splitters \cite{wang2019boson, bayerbach2023bell, Cao2024GHZ}, analysis of rising/falling edge of single\nobreakdash{-}pixel SNSPD signals~\cite{zhu2020resolving, sauer2023resolving, schapeler2023well}, spatial multiplexing of SNSPD arrays in an independent readout configuration~\cite{zhang201916, resta2023gigahertz, craiciu2023high} and our {\ps} architecture. This comparison is summarized in \cref{tab:photon_detection}. 

Starting with TESs, they are known to possess intrinsic PNR capabilities due to their bolometric principle; thus, they satisfy requirements (i) and (a). To guarantee requirement (ii) up to high\nobreakdash{-}photon number events~\cite{morais2024precisely}, post\nobreakdash{-}processing of the electrical pulses must be performed, and, since TESs usually display a slow recovery time, in the order of several hundreds of ns~\cite{morais2024precisely}, this poses a severe limit to their operational speed. Combined with the complex cryogenics needed to reach the mK operating temperature, TESs fail to meet features (b) and (c). 

A common approach to enable PNR capabilities with single-pixel SNSPDs has been to use a plurality of detectors and split the input optical signal, using lossy beam-splitters, amongst them \cite{wang2019boson, bayerbach2023bell, Cao2024GHZ}. In this approach, the requirement (i) cannot easily be met, as the number of multiplexed detectors is limited by the losses of the optical beam splitters and by the increased complexity of the read-out. In this configuration, the requirement (ii) is always met, and the detectors are able to perform PNR discrimination with light pulses of any duration, thus also satisfying (a).  However, each single-pixel detector is still limited to low count rates, and it lacks scalability. Thus, this method fails to meet features (b) and (c).

A different approach relies on the analysis of the time difference in the electrical pulse rising/falling edge of the single-pixel SNSPD signals, which is impacted by the absorption of different photon numbers \cite{zhu2020resolving, sauer2023resolving, schapeler2023well}. Analysis of the trace with a fast oscilloscope can also be used in this approach \cite{sauer2023resolving}. Though requirement (i) is satisfied, it becomes difficult to fulfill (ii): good discrimination between different photon\nobreakdash{-}number events is achieved only up to 3\nobreakdash{-}photon event, becoming more and more blurred with increasing photon\nobreakdash{-}numbers. Moreover, this method requires detectors with extremely low jitter and time tagging devices with ps-level resolution, and it is limited to operation at a few MHz rate, thus making it extremely difficult to accomplish features (b) and (c). Lastly, this method is limited to a few\nobreakdash{-}tens of ps light pulses at most since the different hotspots need to be generated at the same time~\cite{zhu2020resolving, sauer2023resolving, schapeler2023well}, thus failing to meet feature (a). This method can be combined with using multiple detectors with optical beam-splitters, thus providing better PNR performances. It will, however, still suffer from the limitation highlighted above in terms of speed and read-out complexity.  

SNSPD arrays with an independent read-out configuration have also been used as PNR detectors \cite{zhang201916, resta2023gigahertz}. With this implementation, requirement (i) cannot be fully reached; however, increasing the number of pixels can bring this scheme closer and closer to the ideal $n$\nobreakdash{-}photon efficiencies. Requirement (ii) is always satisfied without the need for any post\nobreakdash{-}processing of the electrical pulses, and the devices can work at high-detection rates (feature (b)) and with any light pulse duration (feature (a)). However, the need for 1 coaxial cable per pixel limits the number of pixels that can be made on a single device (e.g., a single device with \textit{n} independent pixels will require an entire \textit{n}-channel cryostat), thus failing to meet the scalability requirement (feature (c)). 

The architecture presented here has a similar limitation compared to arrays with independent pixel read-out in terms of fulfilling requirement (i); however, with this method, almost ideal $n$\nobreakdash{-}photon efficiencies can be reached by increasing the number of pixels without impacting the read-out complexity. The scaling of our parallel technology is ultimately limited by the amount of current being redistributed among the parallel pixels and thus not going into the read-out path. It would be possible, by optimizing the design in terms of resistances and inductances, to increase the number of pixels and further improve the \textit{n-}photon efficiencies. It is however important to consider for which application the PNR detector should be used, as different applications will require discrimination only until a certain photon number state. For example, if we want to use our detector to filter-out multi-photon emissions from an heralded single-photon source \cite{davis2022improved,sempere2022reducing,stasi2023high}, the {\ps} already provides a $C_{>1}$ confidence that is very close to the intrinsic PNR, thus an increase of the number of pixels will not yield a huge improvement in performance. On the contrary, applications such as Gaussian Boson sampling that require high $n$-photon efficiency until higher photon-number states would greatly benefit from an increased number of pixels. Regardless of the application, improving the 1-photon efficiency of the device also remains paramount if high \textit{n-}photon efficiencies are needed. As we showed in \ref{fig:pnr}(b), the assignment probability for the parallel architecture is 1, thus completely fulfilling requirement (ii). Features (a) and (b) are partially fulfilled since the {\ps} can work until a few hundred ps light pulses, and we demonstrate that it can operate until a 40~MHz rate. Finally, it needs only 1 coaxial cable, regardless of the number of pixels that the device has, does not require high\nobreakdash{-}precision time tagging devices, and can discriminate photon numbers in real\nobreakdash{-}time, thus fulfilling (c) as the most scalable solution. It is important to remark that to discriminate up to $n$- photon events, the signal needs to be split into $n$ channels, if using the read-out scheme described here. This could pose some challenges to the room temperature readout and counting electronics, as an increased amount of separate counting channels would be needed. This analysis is summarized in \cref{tab:photon_detection}.\\

\begin{table}[h!]
    \centering
    \small
    \renewcommand{\arraystretch}{1.5} 
    \resizebox{\textwidth}{!}{%
    \begin{tabular}{>{\raggedright\arraybackslash}m{3.0cm} >{\centering\arraybackslash}m{3cm} >{\centering\arraybackslash}m{2.5cm} >{\centering\arraybackslash}m{2.5cm} >{\centering\arraybackslash}m{4.5cm} >{\centering\arraybackslash}m{2cm}}
        \toprule
        & $n$\nobreakdash{-}photon efficiency scaling & Assignment probability $>95\%$ & Operating repetition rate & Readout & Light pulse duration\\
        \midrule
        TES \cite{morais2024precisely} & $\eta^n$ & Up to 14 photon & 100s kHz & SQUID, trace digitalization & 10s of ns\\
        Rising/falling-edge analysis \cite{sauer2023resolving, schapeler2023well} & $\eta^n$ & Up to 3 photon & Few MHz & Time tagging with ps resolution or trace analysis  & 10s of ps \\
        SNSPDs w/ BS \cite{bayerbach2023bell} & $\frac{N!}{(N-n)!}\left(\frac{\eta}{N}\right)^n$ & Yes & 10s MHz & 1 cable per detector and coincidence analysis & Any\\
        Independent read-out arrays \cite{resta2023gigahertz} & $\frac{N!}{(N-n)!}\left(\frac{\eta}{N}\right)^n$ & Yes & 100s MHz & 1 cable per pixel and coincidence analysis & Any\\
        Parallel SNSPD [This work] & $\frac{N!}{(N-n)!}\left(\frac{\eta}{N}\right)^n$ & Yes & 40 MHz & 1 cable and amplitude discrimination & 100s of ps\\
        \bottomrule
    \end{tabular}
    }
    \caption{Performance and operational comparison of different approaches for PNR.}
    \label{tab:photon_detection}
\end{table}

\section{Conclusion}
In conclusion, we present an optimized and scalable parallel SNSPD architecture suitable for diverse applications. We fabricate and characterize a 28-pixel {\ps} detector that, thanks to the parallel architecture, uses a single cryogenic readout channel. The {\ps} shows an 88\% SDE at the single-photon level and is capable of reaching 200~Mcps count rate at 50\% absolute SDE coupled with a timing jitter at 200~Mcps of 80~ps FWHM. Using only four such detectors, with a total of four cryogenic read-out channels, we achieved a $\sim$1.3~Gcps maximum count rate (MCR). The {\ps} also enables a simple approach based on the discrimination of the amplitude of its output voltage to achieve state-of-the-art photon-number capabilities. We report 75\% 2\nobreakdash{-}photon and 62\% 3\nobreakdash{-}photon efficiencies and perform accurate state reconstruction of Poissonian light at a repetition rate of 40~MHz. Finally, we provide an in-depth discussion of different PNR approaches, comparing their strengths and weaknesses. The combination of a scalable read-out, high efficiency, low jitter, and enhanced photon number resolution capabilities at high detection rates makes these detectors well-suited for quantum network applications and high-speed optical quantum computing protocols.

\begin{suppinfo}

Supporting Information. Fabrication process, Thermal-crosstalk analysis, description of measurements apparatus, P-matrix. 

\end{suppinfo}


\section*{Funding Sources}
The authors acknowledge financial support from the Swiss NCCR QSIT, Swiss State Secretariat for Research and Innovation (SERI) (Contract No. UeM019-3), Innosuisse Grant 40740.1 IP-ENG, and from NRC CSTIP Grant QSP043. L.S. is part of the AppQInfo MSCA ITN, which received funding from the EU Horizon 2020 research and innovation program under Marie Sklodowska-Curie Grant Agreement 956071.

\section*{Notes}

\begin{itemize}
\item Conflict of interest/Competing interests: The authors declare no conflicting interests. 
\item Availability of data and materials:
The data that support the findings of this study are available from the corresponding author upon reasonable request.
\item Authors' contributions:
G.V.R., H.Z. and F.B. conceptualized the project. G.V.R, L.S. and T.T. carried out the characterization of the detectors and all the experiments. G.V.R., L.S and T.T. performed the the data analysis and generated the graphs. L.S. and T.T. fabricated the detectors. G.V.R, R.T. and F.B. supervised the project. G.V.R., T.T. and L.S. wrote the manuscript with contributions from all co-authors. 
\end{itemize}

\noindent

\bibliography{references}

\end{document}